\documentstyle[aps,epsfig]{revtex}
\twocolumn[\hsize\textwidth\columnwidth\hsize\csname @twocolumnfalse\endcsname
 
\title{
Thermodynamic Properties of the S=1/2 Heisenberg Chain with
Staggered Dzyaloshinsky-Moriya Interaction}
\author{
Naokazu {\sc Shibata}
and Kazuo {\sc Ueda}${}^*$}
\address{
Department of Basic Science, University of Tokyo,
Komaba, Meguro-ku, Tokyo 153-8902 \\
${}^*$Institute of Solid State Physics, University of Tokyo,
Kashiwa, Chiba 277-8581 and \\
Advanced Science Research Center, JAERI, Tokai, Ibaraki 319-1195}
\maketitle
 
\begin{abstract}
Thermodynamic properties of the S=1/2 Heisenberg chain in
transverse staggered magnetic field $H^y_s$ and
uniform magnetic field $H^x$ perpendicular to the staggered field
is studied by the
finite-temperature density-matrix renormalization-group method.
The uniform and staggered magnetization and specific heat
are calculated from zero temperature to
high temperatures up to $T/J=4$ under various strength of magnetic fields
from $H^y_s/J, H^x/J=0$ to $2.4$.
The specific heat and magnetization of the effective Hamiltonian
of the Yb${}_4$As${}_3$ are also presented, and field induced gap formation and
diverging magnetic susceptibility at low temperature are shown.
\end{abstract}

\begin{document}
\sloppy
%\maketitle
 
%\narrowtext
\section{Introduction}
The one-dimensional spin systems exhibit
many interesting physical phenomena.
Although these systems have only spin degrees of freedom,
their ground state and thermodynamic 
behavior have diversity with various types of magnetic interactions.
Recently, microscopic analysis\cite{Shiba} on Yb${}_4$As${}_3$, 
a typical system showing field induced gap opening at low temperature, 
confirmed that its effective Hamiltonian is a S=1/2 Heisenberg model with 
Dzyaloshinsky-Moriya (DM) interaction. 
This microscopic analysis supports previous scenario for
the field induced gap opening due to the DM 
interaction\cite{Oshi1,Oshi2,Ueda}, 
and stimulates more precise and detailed comparison 
between experiment and theory.

The Yb${}_4$As${}_3$ is a 4$f$ electron system, which undergoes
charge ordering at 290K\cite{Ybc1}. 
Below this temperature one of the four Yb ions becomes trivalent and 
forms one-dimensional (1D) chain along [111] direction\cite{Ybc2,Fulde}.
The Yb${}^{3+}$ ion has one hole in the 4$f$ closed shell. The 
$J=7/2$ ground multiplet splits into four doublets by the 
crystalline field effect.  Thus the low temperature dynamics is 
described by an effective S=1/2 spin chain. 
The neutron scattering experiments on Yb${}_4$As${}_3$
actually confirmed that the excitation spectrum is well described
by the 1D S=1/2 isotropic Heisenberg model\cite{Kho1,Kho2}.

Unusual features which can not be understood from the 
1D isotropic Heisenberg model
are observed under applied magnetic fields. 
One is the upturn of the magnetic susceptibility at low temperature,
and the another is the formation of the excitation gap\cite{Yb1,Yb2}. 

Recently, it was shown that these unusual features are caused 
by the staggered DM interaction, which generates effective staggered 
field under applied magnetic fields\cite{Ueda}.
Since the staggered spin correlation length 
$\xi (q$=$\pi)$ of the Heisenberg model diverges at $T=0$, 
drastic change occurs in the ground state. 
By using a mapping on to the sine-Gordon model, it has been shown 
that the excitation gap is formed by any finite staggered 
field\cite{Oshi1,Oshi2}.
The analytic expressions of the gap,
magnetic susceptibility, and specific heat of Yb${}_4$As${}_3$
are calculated within the sine-Gordon theory, and they are consistent 
with the experiments at low temperatures\cite{Ueda}.

However the sine-Gordon model describes only the low energy physics 
of the effective one-dimensional S=1/2 model. 
In the present paper, we directly calculate thermodynamic 
quantities of the effective Hamiltonian of Yb${}_4$As${}_3$ 
over a wide range of 
temperatures and magnetic fields with high accuracy
extending the limitation of the field theoretical treatment
on the sine-Gordon model.  Such precise calculation over wide 
range of temperatures and magnetic fields is necessary to analyze 
detailed behaviors of Yb${}_4$As${}_3$.

\section{Model and method}

The microscopic study on the Yb${}_4$As${}_3$ shows that its low temperature
effective Hamiltonian is the S=1/2 anisotropic Heisenberg model
with the DM interaction:\cite{Shiba}
\begin{eqnarray}
  {\cal H} &=& J \sum_{i} \left\{S_i^z S_{i+1}^z + \cos{2\theta} 
   ( S_i^x S_{i+1}^x + S_i^y S_{i+1}^y )\right\} \nonumber \\ 
           &+& J \sin{2\theta} \sum_i (-1)^i 
          ({\bf S}_{i} \times {\bf S}_{i+1})^z +  g_\perp H \sum_i S^x_i .
\end{eqnarray}
Here, $H$ is the uniform external magnetic field perpendicular 
to the 1D spin-chain.
The DM interaction is eliminated by rotating 
the spins in the staggered way in the x-y plane;
\begin{equation}
%\left\{
\begin{array}{lllllll}
 {S}^x_i &=& &\cos{\theta}\ \hat{S}^x_i &+& (-1)^i &\sin{\theta}\ \hat{S}^y_i, \\
 {S}^y_i &=& -(-1)^i &\sin{\theta}\ \hat{S}^x_i &+& &\cos{\theta}\ \hat{S}^y_i,\\
{S}^z_i &=&  \hat{S}^z_i .&&&&
\end{array}
%\right.
\end{equation}
After this transformation, the Hamiltonian is mapped on to
\begin{equation}
  {\cal H}= J \sum_{i} {\bf \hat{S}}_{i} \cdot {\bf \hat{S}}_{i+1} + 
 g_\perp H^x \sum_i \hat{S}^x_i +  g_\perp H^y_s \sum_i (-1)^i \hat{S}^y_i ,
\end{equation}
where $ H^x =  H \cos{\theta}$ and $H^y_s =  H \sin{\theta}$.
This is equivalent to the {\it isotropic} Heisenberg model when external
magnetic field is absent. Under finite external magnetic field
$H$, both the effective uniform field $H^x$ and 
transverse staggered field $H^y_s$ are applied to the Heisenberg spin-chain.

In order to calculate thermodynamic quantities, we employ the
finite-temperature density-matrix renormalization-group (finite-$T$ DMRG)
method\cite{FTRG1,FTRG2,FTRG3}.
Since the finite-$T$ DMRG method iteratively expands the quantum transfer
matrix in $\beta$-direction restricting the number of basis states,
the thermodynamic quantities are obtained for desired temperature.
Thermodynamic quantities are obtained from
the maximum eigenvalue and its eigenvector
of the quantum transfer matrix, and the extrapolation
on the system size is not needed\cite{Betsu}.
The numerical error is estimated from the eigenvalues of the density
matrix which are truncated off, and thus the accuracy is systematically
improved by increasing the number of basis states.

In the presence of both the uniform and transverse magnetic fields,
total $S^z$ is no more a conserved quantity and finite matrix elements
are generated between different subspaces of total $S^z$.
However, as will be shown in the following accurate results
can be obtained.
In the present calculation, the truncation error is
smaller than $10^{-4}$ even at the lowest temperature $T/J=0.01$.

The ground state properties are calculated by the standard algorithm
of the DMRG\cite{DMRG} with up to 200 spins with the periodic boundary
conditions.
The thermodynamic limit is taken by using finite size scaling.

\begin{figure}[t]
\epsfxsize=75mm \epsffile{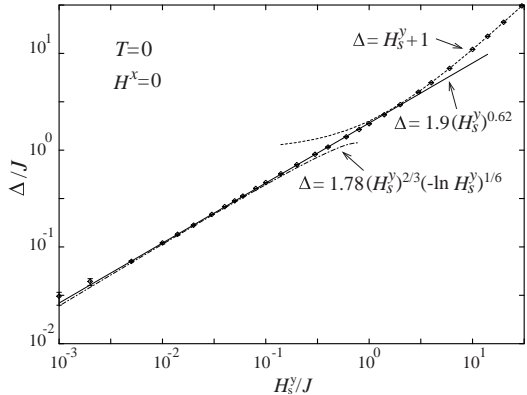}
\caption{
Excitation gap of the 1D Heisenberg model in the staggered field $H^y_s$.
The uniform field $H^x$ is set to be zero. $J=1$. $g_\perp = 1$.}
\end{figure}

\begin{figure}[t]
\epsfxsize=75mm \epsffile{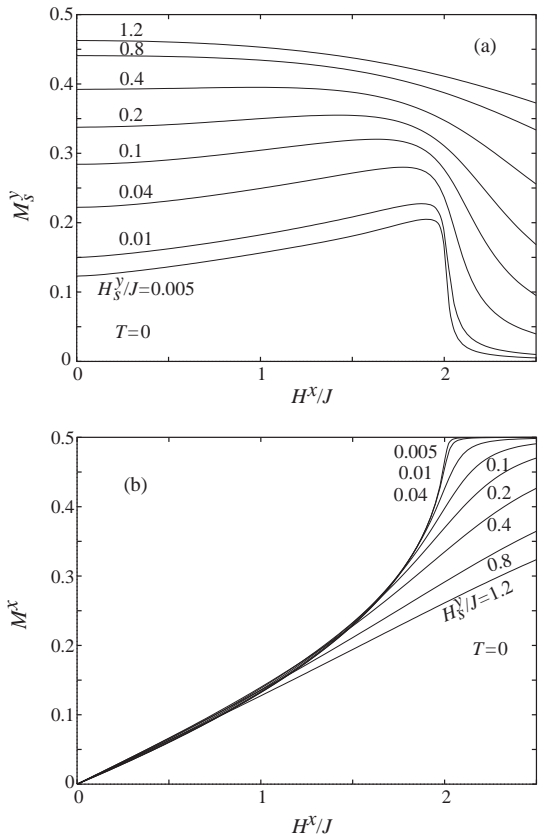}
\caption{
(a) Transverse staggered magnetization,
(b) Uniform magnetization
in the transverse staggered field $H^y_s$ and the
uniform field $H^x$ at $T=0$. $g_\perp = 1$.
}
\end{figure}

\begin{figure}[t]
\epsfxsize=75mm \epsffile{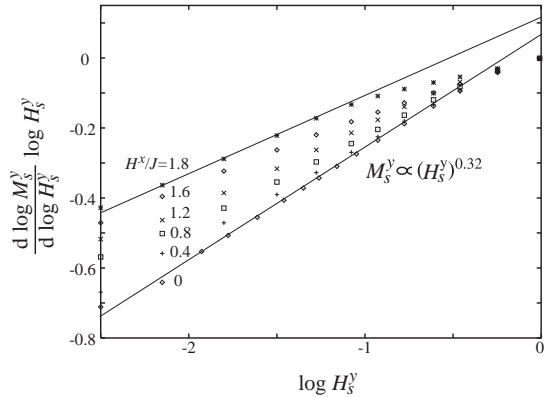}
\caption{
Transverse staggered magnetization at $T=0$. $g_\perp = 1$.
}
\end{figure}
 
\section{Excitation gap and magnetization}
 
In Fig.~1 we show the excitation gap $\Delta$ as a function of
staggered field $H^y_s$.
The gap follows power law in the weak staggered field,
and is well fitted by the exponent 0.62
between $H^y_s\sim 0.002J$ and $2J$.
The power law dependence is also obtained in the field theory,
which predicts $\Delta=1.78(H^y_s)^{2/3}(-\ln{H^y_s})^{1/6}$
including logarithmic correction\cite{Oshi1,Oshi2}.
This result is plotted in the figure by the dashed line. We
find good agreement below $H^y_s\sim 0.1J$.
In the region of intermediate strength of the staggered field
$H^y_s\sim J$, the gap slightly deviate from the predicted power law,
but still is quite close to the value $\Delta=1.9(H^y_s)^{0.62}$.
For strong field  $H_s^y > 2J$,
the spins are almost completely aligned to the staggered field
and the gap is given by $\Delta= H_s^y+J$ as shown in the dotted
line.

The magnetization induced by the magnetic field is shown in Fig.~2.
The transverse staggered magnetization $M^y_s$ at $T=0$ is shown in 
Fig.~2 (a), and the uniform magnetization $M^x$ is shown in Fig.~2 (b). 
For small staggered fields, the staggered magnetization monotonically 
increases with increasing uniform field until $H^x\sim 2J$.
This result shows that the uniform field enhances the transverse staggered
susceptibility. At $H^x\sim 2J$, however, the staggered magnetization
sharply decreases. The decrease in $M^y_s$ is related to the sharp
increase in uniform magnetization $M^x$ as shown in Fig.~2 (b).
The substantial change in the ground state at $H^x=2J$ is due to the
singular behavior at the saturation of $M^x$ in the limit of
weak $M^y_s$.
For large $H^y_s$, both the staggered magnetization and
the uniform magnetization show monotonic change and
no singular behavior is observed at $H^y\sim 2J$.

\begin{figure}[t]
\epsfxsize=75mm \epsffile{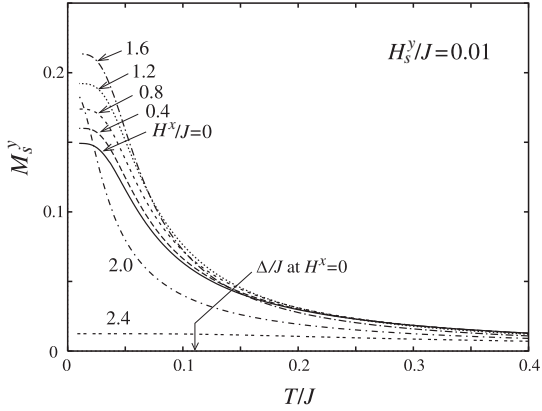}
\caption{
Temperature dependence of transverse staggered magnetization
at $H^y_s=0.01J$. $g_\perp = 1$. The excitation gap $\Delta$ is $0.11J$ at $H^x=0$.
}
\end{figure}
 
\begin{figure}[t]
\epsfxsize=75mm \epsffile{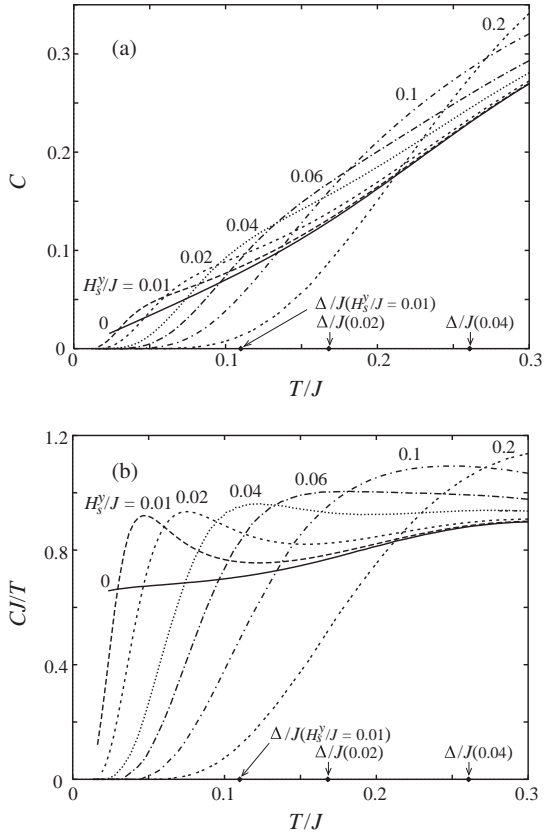}
\caption{
Specific heat in the staggered field $H^y_s$. $H^x=0$. $g_\perp = 1$.
The excitation gap $\Delta$ is shown on the horizontal axis.
}
\end{figure}

In Fig.~3 we show the staggered field dependence of 
the staggered magnetization at fixed uniform field. 
We plot
\[
\frac{d \log{M^y_s}}{d \log{H^y_s}} \log{H^y_s}
\]
to eliminate logarithmic dependence on the staggered field.
In this figure we find linear dependence for weak
staggered field. This result shows that $M^y_s$ depends
on $H^y_s$ as $M^y_s\propto (H^y_s)^\alpha (\log{H^y_s})^\beta$.
The power law exponent $\alpha$ is determined by the slope in the 
figure, and $\alpha=0.32$ is obtained at $H^x=0$.
This is almost consistent with the result 1/3 obtained by the field
theory \cite{Oshi1,Oshi2}.
With increasing uniform field up to $H^x\sim 1.8J$,
the power law exponent $\alpha$ decreases down to 0.24.
The field theory also predicts the lowering of the power law
exponent\cite{Oshi1,Oshi2}.

The temperature dependence of the staggered magnetization at 
$H^y_s=0.01J$ is shown in Fig.~4.
At high temperature, $M^y_s$ is almost the same for
$H^x\stackrel{<}{_\sim} 1.6J$. 
With decreasing temperature, $M^y_s$ increases monotonically
and becomes almost constant below $T/J\sim 0.2\Delta$.
For $H^x=2J$, the staggered magnetization is small compared with that
for small uniform fields $H^x\stackrel{<}{_\sim} 1.6J$, 
but it continues to increase 
even at low temperatures below $T/J\sim 0.01$ due to the strongly 
reduced gap at $H^x=2J$.
For $H^x>2J$, the spins are almost completely aligned to the
uniform magnetic field and the temperature dependence is small.

\begin{figure}[t]
\epsfxsize=75mm \epsffile{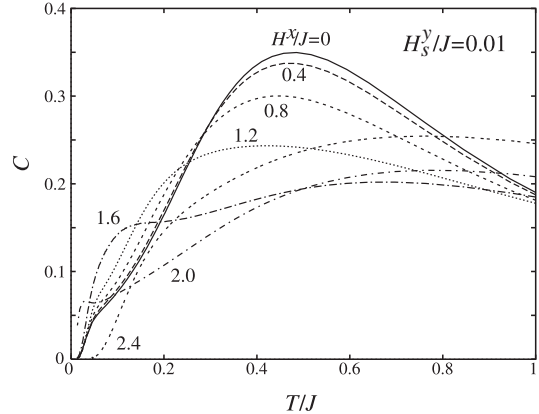}
\caption{
Specific heat in the transverse staggered field $H^y_s$
and the uniform field $H^x$. $g_\perp = 1$.
}
\end{figure}

\section{Specific heat}
Fig.~5 shows the specific heat at several strength of the staggered 
field $H^y_s$ under zero uniform field $H^x=0$. Since the excitation gap 
exists for finite $H^y_s$, deviation occurs from the $T$-linear 
dependence at $T \sim \Delta$. The specific heat has a broad 
peak below $T \sim \Delta$, and then exponentially decreases down 
to zero.
As shown In Fig.~5 (b), which show the specific heat divided by temperature,
the sharp decrease occurs around the temperature
$T \sim 0.2\Delta$ with a broad peak at higher temperature $T \sim 0.4\Delta$. 
The exponential dependence due to the gap is observed only below 
$T \sim 0.2\Delta$.
 
The effects of uniform field $H^x$ under a finite staggered 
field $H^y_s=0.01J$ is shown in Fig.~6.
The low temperature behavior below $T \sim 0.2\Delta$
is almost the same for $H^x \stackrel{<}{_\sim}  1.6J$.
Since the low temperature behavior is characterized by the gap,
this result shows almost constant gap for 
$H^x \stackrel{<}{_\sim} 1.6J$.
In contrast to the behavior below $T \sim 0.2\Delta$, the broad peak
at intermediate temperature strongly depend on the uniform field.
The maximum of the specific heat around $T/J \sim 0.4$ at $H^x=0$
becomes smaller and shifts to lower temperature with increasing $H^x$.
At $H^x=1.6$ the small dip appears at $T/J\sim 0.2$, and the main
peak is divided into two peaks.
With increasing $H^x$, the peak at lower temperature shifts toward $T=0$.
The low temperature behavior is strongly modified at $H^x=2J$, and 
the peak at low temperature disappears for $H^x>2J$.
With further increasing $H^x$, the gap turns to increase with
clear exponential temperature dependence at low temperature.

\section{Application to Yb${}_4$As${}_3$}

The magnetization of Yb${}_4$As${}_3$ is calculated from
the effective Hamiltonian Eq.~(3) with rotating the spins 
back to the original coordinates. Thus the magnetization
perpendicular to the 1D chain of Yb${}_4$As${}_3$ 
is given by $M=M^x \cos \theta + M^y_s \sin \theta$.
The calculated $M$ are shown in Fig.~7 for various $\tan\theta$, 
which is related to the ratio between the DM interaction ($D_{DM}$) and $J$
as $\sin{2\theta}=D_{DM}/J$. 
The magnetization sharply increases
near $H=0$, and the exponent is 0.32 in the limit of $H=0$.
The magnetization curves gradually change its shape
with decreasing $\tan{\theta}$.
By fitting the present results with the experimentally obtained 
magnetization using the experimentally estimated value of 
$g_\perp =1.3$, $\tan{\theta}$ is estimated to be 0.19\cite{Iwasa}.

Temperature dependence of the magnetization divided by the 
magnetic field $H$ is presented in Fig.~8. In this calculation we
use the parameters $\tan{\theta}=0.19$, $J=26K$, and $g_\perp =1.3$
which reasonably reproduce susceptibility data\cite{Iwasa}. 
We also plot the susceptibility of 1D Heisenberg 
antiferromagnet (1D-HAF) for comparison. 
At high temperatures $T>J$ the susceptibility is almost the
same to that of 1D-HAF. With decreasing 
temperature, however, the susceptibility of Yb${}_4$As${}_3$ 
largely increases especially for small magnetic field
due to the component of the transverse staggered susceptibility 
of the effective Heisenberg model\cite{Ueda}.
The susceptibility diverges at $T=0$ in the limit of $H=0$.
This is caused by the fact that the staggered correlation length of the 
Heisenberg model $\xi (q$=$\pi)$ diverges at $T=0$.

%\section{Specific heat of Yb${}_4$As${}_3$}
The specific heat of Yb${}_4$As${}_3$ under the magnetic field 
perpendicular to the 1D chain is presented in Fig.~9. The 
temperature dependence is similar to the results
in Fig.5, but in the present case, uniform field about four times
larger than the effective staggered field induced by the DM 
interaction is present.
Similarly to the results in Fig.5 the specific heat $C/T$
has a maximum around $T= 0.4\Delta \sim 0.6\Delta$ and 
exponentially decreases down to zero.

It is highly desired that the theoretical results presented 
in this paper will be compared with experimental results for
a monodomain sample under magnetic fields perpendicular to 
the chain direction.

\begin{figure}[t]
\epsfxsize=75mm \epsffile{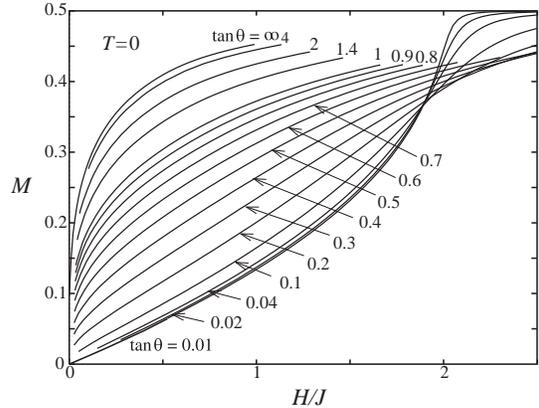}
\caption{
Magnetization for various $\tan\theta$ at $T=0$. $g_\perp = 1$.
}
\end{figure}
 
\begin{figure}[t]
\epsfxsize=75mm \epsffile{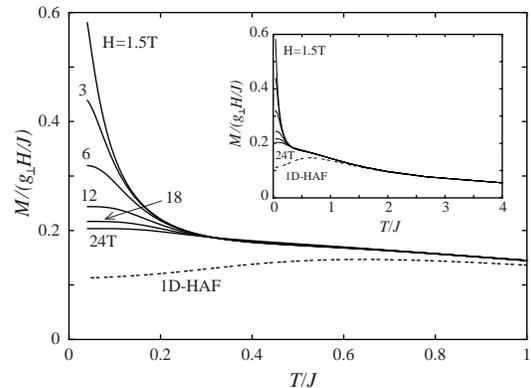}
\caption{
Temperature dependence of the susceptibility of the effective Hamiltonian of 
Yb${}_4$As${}_3$. The magnetic field is applied perpendicular to the 1D chain. 
$\tan\theta=0.19$, $J=26K$, $g_\perp =1.3$. 1D-HAF is the susceptibility of 
1D Heisenberg antiferromagnetic chain in the limit of $H\rightarrow 0$.
}
\end{figure}

\begin{figure}
\epsfxsize=75mm \epsffile{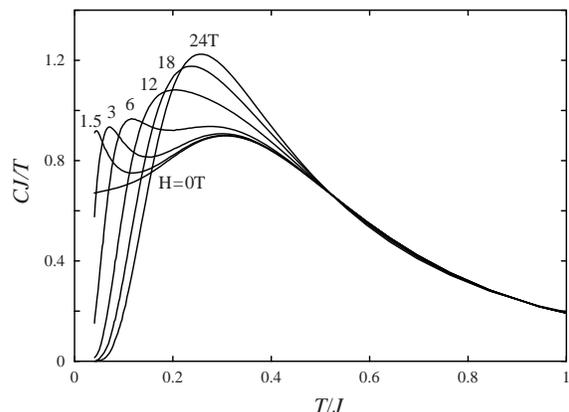}
\caption{
Specific heat of the effective Hamiltonian of Yb${}_4$As${}_3$
in magnetic field perpendicular to the 1D chain. 
$\tan\theta=0.19$, $J=26K$, $g_\perp =1.3$. 
}
\end{figure}

\section*{Acknowledgement}
We acknowledge Dr. K. Iwasa, Prof. A. Ochiai and Dr. H. Aoki for 
valuable discussions on the experimental results of magnetization 
and specific heat of Yb${}_4$As${}_3$.


\begin{thebibliography}{99}
\addcontentsline{toc}{section}{References}
%\bibitem{BA}M. Takahashi: Prog. Theor. Phys. {\bf 47} (1972) 69;
%ibid. {\bf 52} (1974) 103.
\bibitem{Shiba}H. Shiba, K. Ueda, and O. Sakai: J. Phys. Soc. Jpn. {\bf 69}
(2000) 1493.
\bibitem{Oshi1}M. Oshikawa and I. Affleck: Phys. Rev. Lett. {\bf 79} (1997) 
2883.
\bibitem{Oshi2}I. Affleck and M. Oshikawa: Phys. Rev. B {\bf 60} (1999) 1038.
\bibitem{Ueda}M. Oshikawa, K. Ueda, H. Aoki, A. Ochiai and M. Kohgi:
J. Phys. Soc. Jpn. {\bf 68} (1999) 3181.
\bibitem{Ybc1}A. Ochiai, T. Suzuki and T. Kasuya:
J. Phys. Soc. Jpn. {\bf 59} (1990) 4129.
\bibitem{Ybc2}M. Kohgi, K. Iwasa, A. Ochiai, T. Suzuki, J.-M. Mignot,
B. Gillon, A. Gukasov, J. Schweizer, K. Kakurai, M. Nishi, A. Donni and
T. Osakabe: Physica B {\bf 230-232} (1997) 638.
\bibitem{Fulde}P. Fulde, B. Schmidt and P. Thalmeier: Europhys. 
Lett. {\bf 31} (1995) 323.
\bibitem{Kho1}M. Kohgi, K. Iwasa, J.-M. Mignot, A. Ochiai and T. Suzuki:
Phys. Rev. B {\bf 56} (1997) R11388.
\bibitem{Kho2}M. Kohgi, K. Iwasa, J.-M. Mignot, N. Pyka, A. Ochiai, H. Aoki
and T. Suzuki: Physica B {\bf 259-261} (1999) 269.
%\bibitem{Cu}D. C. Dender, P. R. Hammar, Daniel H. Reich, C. Broholm,
%and G. Aeppli: Phys. Rev. Lett. {\bf 79} (1997) 1750.
\bibitem{Yb1}P. H. P. Reinders, U. Ahlheim, K. Fraas, F. Steglich and
T. Suzuki: Physica B {\bf 186-188} (1993) 434.
\bibitem{Yb2}R. Helfrich, M. K\"oppen, M. Lang, F. Steglich and A. Ochiai:
J. Magn. Magn. Mater. {\bf 177-181} (1998) 309.
\bibitem{Betsu}H. Betsuyaku: Phys. Rev. Lett. {\bf 53} (1984) 629;
Prog. Theor. Phys. {\bf 73} (1985) 319.
\bibitem{FTRG1}R. J. Bursill, T. Xiang and G. A. Gehring: J. Phys.:
Condens. Matter. {\bf 8} (1996) L583.
\bibitem{FTRG2}X. Wang and T. Xiang: Phys. Rev. B {\bf 56} (1997) 5061.
\bibitem{FTRG3}N. Shibata: J. Phys. Soc. Jpn. {\bf 66} (1997) 2221.
\bibitem{DMRG}S. R. White: Phys. Rev. Lett. {\bf 69} (1992) 2863;
Phys. Rev. B {\bf 48} (1993) 10345.
\bibitem{Iwasa}K. Iwasa, M. Kohgi, A Gukasov, J.-M. Mignot, 
N. Shibata, A. Ochiai, H. Aoki, and T. Suzuki:
to be published.
\end{thebibliography}
\end{document}